\documentclass[12pt]{article}

\usepackage{scicite}

\usepackage{times}

\usepackage{graphicx}

\newenvironment{sciabstract}{%
\begin{quote} \bf}
{\end{quote}}

\renewcommand\refname{References and Notes}

\newcounter{lastnote}

\title{Detection of circumstellar material in a normal Type Ia Supernova}

\author
{ F. Patat$^{1\ast}$, P. Chandra$^{2,3}$, R. Chevalier$^{2}$,
S. Justham$^{4}$, Ph.\ Podsiadlowski$^{4}$, C. Wolf$^{4}$,\\
A. Gal-Yam$^{5}$,  L. Pasquini$^{1}$, I.A. Crawford$^{6}$,
P.A. Mazzali$^{7,8}$, A.W.A. Pauldrach$^{9}$\\
K. Nomoto$^{10}$, S. Benetti$^{11}$, E. Cappellaro$^{11}$,
N. Elias-Rosa$^{7,12}$,  W. Hillebrandt$^{7}$,\\ D.C. Leonard$^{13}$, A.
Pastorello$^{14}$, A. Renzini$^{11}$, F. Sabbadin$^{11}$,
J.D. Simon$^{5}$, M. Turatto$^{11}$
\\
\\
\small{$^{1}$European Southern Observatory, K. Schwarzschild Str.2, 
	85748, Garching b. M\"unchen, Germany}\\
\small{$^{2}$University of Virginia, Dept. of Astronomy, PO Box 400325,
Charlottesville, VA 22904, USA}\\
\small{$^{3}$ Jansky Fellow, National Radio Astronomy Observatory}\\
\small{$^{4}$Dept. of Astrophysics, University of Oxford, Oxford OX1 3RH, UK}\\
\small{$^{5}$Astronomy Department, MS 105-24, California Institute of
Technology, Pasadena, CA 91125,USA}\\
\small{$^{6}$School of Earth Sciences, Birkbeck College London,
Malet Street, London WC1E 7HX, UK}\\
\small{$^{7}$Max-Planck-Institut f\"ur Astrophysik, K. Schwarzschild Str.1,
85748, Garching b. Muenchen, Germany}\\
\small{$^{8}$INAF-Osservatorio Astronomico, v. Tiepolo 11, 34131 Trieste, Italy}\\
\small{$^{9}$Institut f\"ur Astronomie und Astrophysik der Ludwig-Maximilians-Universit\"at, 81679 Munich, Germany}\\
\small{$^{10}$Dept. of Astronomy, University of Tokyo, Bunkyo-ku, Tokyo 113-0033, Japan}\\
\small{$^{11}$INAF-Osservatorio Astronomico, v. Osservatorio 5,
35122 Padova, Italy}\\
\small{$^{12}$Universidad de La Laguna, Av. Astrof\'isico Fransisco 
S\'anchez s/n, E-38206, La Laguna, Tenerife, Spain}\\ 
\small{$^{13}$Department of Astronomy, San Diego State University, San
Diego, California 92182, USA}\\
\small{$^{14}$Astrophysics Research Centre, Queen's University Belfast, BT7 1NN, UK}\\
\small{$^\ast$To whom correspondence should be addressed. E-mail: fpatat@eso.org}
}

\date{}

\begin{document} 


\baselineskip24pt


\maketitle 

\vspace{10mm}

\begin{center}
{\sc Accepted for publication in Science}

(Submitted on March 26, 2007 - Accepted on June 29, 2007)

\end{center}

\newpage


\begin{sciabstract}
Type Ia supernovae are thought to be thermonuclear explosions of
accreting white dwarfs that reach a critical mass limit. Despite their
importance as cosmological distance indicators, the nature of their
progenitors has remained controversial. Here we report the detection
of circumstellar material in a normal Type Ia supernova. The expansion
velocities, densities and dimensions of the circumstellar envelope
indicate that this material was ejected from the progenitor
system. The relatively low expansion velocities appear to favor a
progenitor system where a white dwarf accretes material from a
companion star which is in the red-giant phase at the time of
explosion.
\end{sciabstract}


Due to their extreme luminosities and high homogeneity, Type Ia
Supernovae have been used extensively as cosmological reference
beacons to trace the evolution of the Universe {\it(1, 2)}. However,
despite significant recent progress, the nature of the progenitor
stars and the physics which govern these powerful explosions have
remained very poorly understood {\it(3, 4)}.  In the presently
favored single-degenerate model, the supernova progenitor is a white
dwarf in a close binary accreting from a non-degenerate companion
{\it(5)}; the white dwarf explodes in a thermonuclear explosion when
it approaches the Chandrasekhar limit.  A direct method for
investigating the nature of the progenitor systems of Type Ia
supernovae (hereafter SNe Ia) is to search for signatures of the
material transferred to the accreting white dwarf in the circumstellar
material (CSM). Previous attempts have aimed at detecting the
radiation which would arise from the interaction between the fast
moving SN ejecta and the slow moving CSM in the form of narrow
emission lines {\it(6)}, radio {\it(7)} and X-ray emission
{\it(8)}. The most stringent upper limit to the mass loss rate set by
radio observations is as low as 3$\times$10$^{-8}$ solar masses per
year ($M_\odot$ yr$^{-1}$) for an assumed wind velocity of
10\,km\,s$^{-1}$ {\it(7)}. Two remarkable exceptions are represented
by two peculiar SNe Ia, SN 2002ic and SN 2005gj, which have shown
extremely pronounced hydrogen emission lines {\it(9, 10)}, that have
been interpreted as a sign of strong ejecta-CSM interaction
{\it(11)}.  However, the classification of these supernovae as SNe Ia
has recently been questioned {\it(12)}, and even if they were SN Ia,
they are unlikely to account for normal Type Ia explosions {\it(7)}
that, so far, lack any signature of mass transfer from a hypothetical
donor.  Here, we report direct evidence of CSM in a SN Ia that has
shown a normal behavior at X-ray, optical and radio wavelengths.

\begin{figure}
\centerline{
\includegraphics[width=140mm]{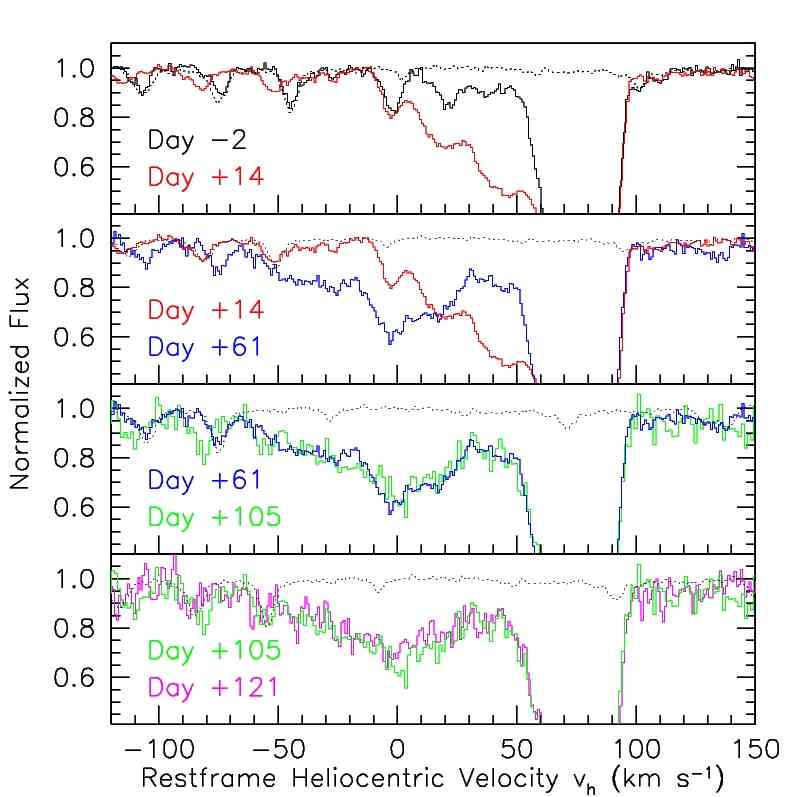}
}
\caption{Time evolution of the Sodium D$_2$ component 
region as a function of elapsed time since $B$-band maximum light. The
heliocentric velocities have been corrected to the rest-frame using
the host galaxy recession velocity. All spectra have been normalized
to their continuum. In each panel, the dotted curve traces the
atmospheric absorption spectrum.}
\end{figure}

SN~2006X was discovered in the Virgo Cluster spiral galaxy NGC~4321
{\it(13)}. A few days later, the object was classified as a
normal Type Ia event occurring 1\,--\,2 weeks before maximum light, which was
affected by substantial extinction {\it(14)}. Prompt Very Large
Array (VLA) observations have shown no radio source at the SN position
{\it(15)}, establishing one of the deepest and earliest limits
for radio emission from a Type Ia, and implying a mass-loss rate of
less than a few 10$^{-8}$ M$_\odot$ yr$^{-1}$ (for a low wind velocity
of 10\,km\,s$^{-1}$). The SN was not visible in the 0.2--10 keV X-rays
band down to the SWIFT satellite detection limit {\it(8)}.

We have observed SN~2006X with the Ultraviolet and Visual Echelle
Spectrograph mounted at the European Southern Observatory 8.2m Very
Large Telescope. Observations were carried out on four different
epochs, which correspond to days $-$2, +14, +61 and +121 with respect
to $B$-band maximum light.  Additionally, a fifth epoch (day +105) was
covered with the High Resolution Echelle Spectrometer mounted at the
10m Keck telescope {\it(16)}. The most remarkable finding from our
data is the clear evolution seen in the profile of the Na~I D lines
(5889.95, 5895.92 \AA). In fact, besides a strongly saturated and
constant component, arising in the host galaxy disk (see section S2,
Fig.~S1), a number of features spanning a velocity range of about 100
km s$^{-1}$ appear to vary significantly with time (Fig.~1, Fig.~S2).
SN~2006X is projected onto the receding side of the galaxy, and the
component of the rotation velocity along the line of sight at the
apparent SN location is about +75 km s$^{-1}$ {\it(17)}, which
coincides with the strongly saturated Na~I D component, the saturated
Ca~II H\&K lines, and a weakly saturated CN vibrational band (0-0)
(Fig.~2, Fig.~S1).  This and the lack of time evolution proves that
the deep absorption arises within the disk of NGC~4321 in an
interstellar molecular cloud (or system of clouds) that is responsible
for the bulk of the reddening suffered by SN~2006X (see text S2).

In contrast, the relatively blue-shifted structures of the Na~I D
lines show a rather complex evolution. The number of features, their
intensity and width are difficult to establish. Nevertheless, for the
sake of discussion, four main components, which we will indicate as A,
B, C and D, can be tentatively identified in the first two epochs
(Fig.~2). Components B, C and D strengthen between day
$-$2 and day +14 while component A remains constant during this time
interval.  The situation becomes more complicated on day +61:
components C and D clearly start to decrease in strength, but
component B remains almost constant and component A becomes definitely
deeper and is accompanied by a wide absorption that extends down to a
rest-frame heliocentric velocity $v_h\simeq -$50 km s$^{-1}$ (Fig.~1,
Fig.~S2).  After this epoch there is no evidence of evolution, and
component A remains the most intense feature up to the last phase
covered by our observations, more than four months after the
explosion.

Variable interstellar absorption on comparably short timescales has
been claimed for some Gamma Ray Bursts (GRB), and it has been
attributed by some authors to line-of-sight geometrical effects, due
to the fast GRB expansion coupled to the patchy nature of the
intervening absorbing clouds {\it(18)}. Our data clearly show that
despite the marked evolution in the Na~I D lines, Ca~II H\&K
components do not change with time (see Fig.~2, Fig.~S3 and the
discussion in sections S3 and S4). Therefore, in the case of SN~2006X,
transverse motions in the absorbing material and line-of-sight effects
due to the fast SN photosphere expansion (typically 10$^4$ km
s$^{-1}$) can be definitely excluded, since they would cause
variations in all absorption features.

For this reason we conclude that the Na~I features seen in SN~2006X,
arising in a number of expanding shells (or clumps), evolve because of
changes in the CSM ionization conditions induced by the variable SN
radiation field. In this context, the different behavior seen in the
Na~I and Ca~II lines is explained in terms of the lower ionization
potential of Na~I (5.1 eV, corresponding to 2417\AA) with respect to
Ca~II (11.9 eV, corresponding to 1045\AA), their different
recombination coefficients and photoionization cross sections coupled
to a UV-deficient radiation field (see text S4). Regrettably, not much
is known about the UV emission of SNe Ia shortwards of 1100\AA\/
{\it(8, 19)}. From a theoretical point of view, a severe UV line
blocking by heavy elements like Fe, Co and Mg is expected
{\it(20)}. An estimate of the Na~I ionizing flux, $S_{UV}$, can be
derived from a synthetic spectrum of a Type Ia SN at maximum light
{\it(21)}, and this turns out to be $S_{UV}\sim$5$\times$10$^{50}$
photons s$^{-1}$. One can verify that this flux is largely sufficient
to fully ionize Na~I up to rather large distances
$r_i\sim$5$\times$10$^{18}$ cm.

Nevertheless, since the recombination timescale $\tau_{r}$ must be of
the order of 10 days, this requires an electron density $n_e$ as large
as 10$^5$ cm$^{-3}$ ( see section S4). Given the low abundance of any
other element, such a high electron density can be produced only by
partial hydrogen ionization. Due to the severe line blocking suffered
by Type Ia SNe {\it(20)}, the flux of photons capable of ionizing H is
very small ($\sim$4$\times$10$^{44}$ photons s$^{-1}$) and this
imposes that the gas where the Na~I time-dependent absorptions arise
must be confined within a few 10$^{16}$ cm from the SN (see text
S4). In a SN of this type, the flux in the 1120--2640\,\AA\/ band
decreases by a factor of ten in the first two weeks after maximum
light {\it(8, 19)}. Since at a distance of $\sim$10$^{16}$ cm from the
SN the ionization timescale $\tau_i$ for Na~I is much shorter than
$\tau_r$, the ionization fraction grows with time following the
increase of the UV flux during the pre-maximum phase, while after
maximum it decreases following the recombination timescale. This would
explain the overall growth of the blue components' depth, as shown by
our data, in terms of an increasing fraction of neutral Na, while the
different evolution of individual components would be dictated by
differences in the densities and distances from the SN.  Moreover,
once all the Na~II has recombined (which should happen within a few
$\tau_r$, i.e. $\sim$1 month), there should be no further evolution,
in qualitative agreement with the observations. Additionally, since
the flux of photons that can ionize Ca~II is more than four orders of
magnitude less than in the case of Na~I (see section S4), the corresponding
ionization fraction is expected to be of a few per cent
only. Therefore, the recombination of Ca~III to Ca~II does not produce
measurable effects on the depth of Ca~II H\&K lines, as it is indeed
observed (see text S3).

\begin{figure}
\centerline{
\includegraphics[width=140mm]{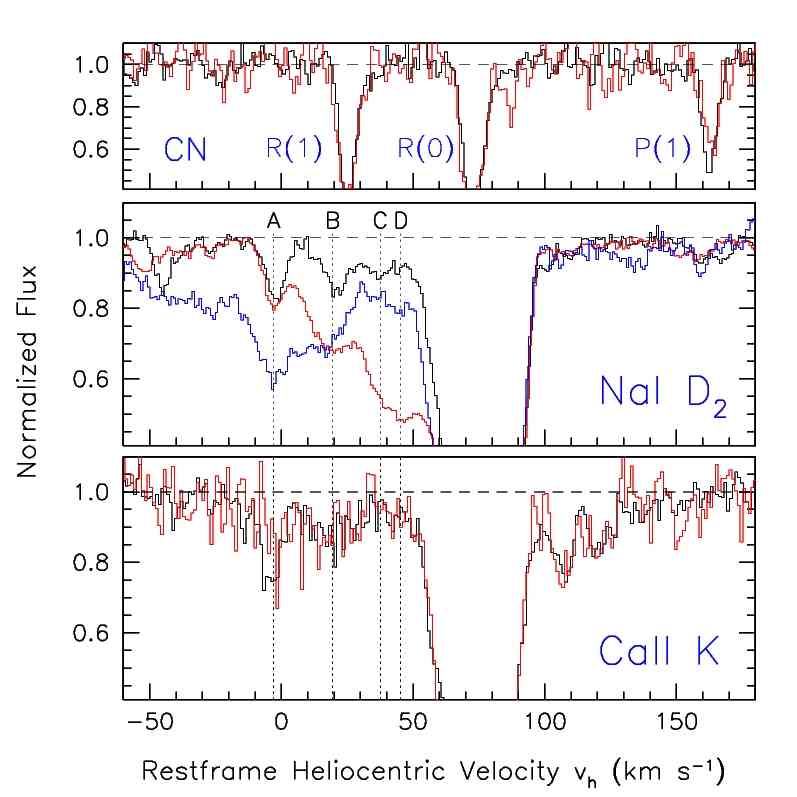}
}
\caption{Evolution of the Na~I D$_2$ and Ca~II K line 
profiles between day $-$2 (black), day +14 (red) and day +61 (blue,
Na~I D$_2$ only). The vertical dotted lines mark the four main variable
components at $-$3 (A), +20 (B), +38 (C) and +45 (D) km s$^{-1}$. For
comparison, the upper panel shows the $R(0)$, $R(1$) and $P(1)$ line
profiles of the (0-0) vibrational band of the CN
$B^2\Sigma-X^2\Sigma$. The velocity scale refers to the $R(0)$
transition (3874.608 \AA). Color coding is as for the other two
panels.}
\end{figure}

The H mass ($M({\rm H\/})$) contained in the shells generating the
observed absorptions can be estimated from our observations after
making some conservative assumptions. The Na~I column density
$N({\rm{NaI\/}})$ deduced from the most intense feature (D, day +14)
is $N({\rm NaI\/})\simeq 10^{12}\; {\rm cm} ^{-2}$. Assuming that the
material generating this component is homogeneously distributed in a
thin spherical shell with radius $10^{17}\,{\rm cm}$, a solar Na/H
ratio ($\log$ Na/H=$-$6.3) and complete Na recombination, an upper
limit to the shell mass can be estimated as $M({\rm H\/})\leq 3 \times
10 ^{-4}\;M_\odot$ (this value is reduced by a factor 100 for material
at about 10$^{16}$ cm, the most likely distance for components C and
D; see below). Even in the case of complete ionization, such a H mass
would produce an H$\alpha$ luminosity of $\sim$4$\times$10$^{34}$ erg
s$^{-1}$, which is two orders of magnitude below the 3-sigma upper
limits set by our observations at all epochs (Table S2) and by any
other SN Ia observed so far {\it(6)}.  Therefore, the absence of
narrow emission lines above the detection limit is not in
contradiction with the presence of partially ionized H up to masses of
the order of 0.01 M$_\odot$.

However, photo-ionization alone cannot account for the fact that not
all features increase in depth with time (Fig.~2). In fact, on day
+61, components C and D turn back to the same low intensity they had
on day $-$2. One possible explanation is that the gas is re-ionized by
some other mechanism, like the ejecta-CSM interaction. In this case,
the absorbing material generating components C and D must be close
enough to the SN so that the ejecta can reach it in about one month
after the explosion ($\sim$10$^{16}$ cm for maximum ejecta velocities
of 4$\times$10$^4$ km s$^{-1}$).  Similarly, in order not to be
reached by the ejecta more than four months after, A, B and the broad
high-velocity components must arise at larger distances
($>$5$\times$10$^{16}$ cm). This scenario is not ruled out by the lack
of radio emission from SN~2006X {\it(15)}. In fact, in the light of
our current understanding of the ejecta-CSM interaction mechanism
{\it(22)}, the presence of similar shells with masses smaller than a
few 10$^{-4}$ M$_{\odot}$, cannot be excluded by radio non-detections
of SNe Ia in general {\it(7)}. Our findings are consistent with upper
limits on the radio flux set by our VLA observations, obtained about
10 months after the explosion (see section S1), which are comparable
to the best upper limits set on the radio luminosity of other normal
SNe Ia {\it(7)}.

If we adopt the velocity of the CN lines as indicative of the host
galaxy rotation component along the line of sight at the SN location,
then our observations provide solid evidence of CSM expanding at
velocities that span a range of about 100 km s$^{-1}$
(Fig.~2).

The most important implication of these observations is that they show
that this circumstellar material was ejected from the progenitor
system in the recent past. For instance, with a shell radius of
10$^{16}$ cm and a wind velocity of $\sim$50 km s$^{-1}$, the material
would have been ejected some 50 years before the explosion.  This
almost certainly rules out a double-degenerate scenario for SN~2006X,
where the supernova would have been triggered by the merger of two CO
white dwarfs. In this case, no significant mass loss would be expected
in the phase immediately preceding the supernova. Thus, a
single-degenerate model is the favored model for SN~2006X, where the
progenitor accreted from a non-degenerate companion star.  

Mean velocities for the circumstellar material of $\sim$50 km s$^{-1}$
are comparable to those reported for the winds of early red giant (RG)
stars {\it(22)}; velocities matching our observations are also
expected for late subgiants. The observed material is moving more
slowly than would be expected for winds from main sequence donor stars
or from compact helium stars. Of the two major formation channels
proposed for SN Ia with a non-degenerate donor star {\it(23)}, these
wind velocities seem more consistent with the shorter-period end of
the "symbiotic" formation channel. The observed structure of the
circumstellar material could be due to variability in the wind from
the companion RG; considerable variability of RG mass loss is
generally expected {\it(24)}.

An alternative interpretation of these distinct features is that they
arise in the remnant shells of successive novae, which can create
dense shells in the slow moving material released by the companion
{\it{(25, 26)}}. This seems to require an aspherical shell geometry in
order to match the observed low velocities (see S6 for further
details). Not only might this be expected a priori {\it{(27)}},
observations of the 2006 outburst of RS Ophiuchi show that there is an
equatorial density enhancement which strongly restrains the expansion
of the nova shell {\it{(28, 29, 30)}}.

One crucial issue is whether what we have seen in SN~2006X represents
the rule or is rather an exceptional case. Other cases of SNe Ia
showing negative velocity components are known, like SN~1991T and
SN~1998es (see Fig.~S5 and the discussion in S5). Unfortunately,
multi-epoch high-resolution spectroscopy is not available for these
objects (to our knowledge, the SN~2006X data set is unique in this
respect), and therefore time variability cannot be
demonstrated. Nevertheless, the data clearly show components
approaching the observer at velocities which reach at least 50 km
s$^{-1}$ with respect to the deep absorption that we infer to be
produced within the disks of the respective host galaxies. This, and
the fact, that SN~2006X has shown no optical, UV and radio peculiarity
whatsoever, supports the conclusion that what we have witnessed for
this object is common to normal SN Ia, possibly all of them, even
though variations due to different inclinations of the line of sight
with respect to the orbital plane may exist.

\footnotesize

\vspace{10mm}

{\bf References and Notes}

\begin{enumerate}
\item[1.]\label{riess98} A.G. Riess {\it et al.}, 
	{\it Astron. J.} {\bf 116}, 1009 (1998).
\item[2.]\label{perlmutter} S. Perlmutter {\it et al.},
	{\it Astrophys. J.\/} {\bf 517}, 565 (1999).
\item[3.]\label{branch} D. Branch, M. Livio, L.R. Yungelson, 
	F. Boffi, E. Baron, {\it Publ. Astr. Soc. Pac.\/}  
	{\bf 107}, 1019 (1995).
\item[4.]\label{hillebrandt} W. Hillebrandt, J.C. Niemeyer, 
	{\it Annu. Rev. Astron. Astrophys.} {\bf 38}, 191 (2000).
\item[5.]\label{whelan} J. Whelan, I. Iben, {\it Astrophys. J.\/} 
	{\bf 186}, 1007 (1973).
\item[6.]\label{mattila} S. Mattila {\it et al.},  
	{\it Astron. Astrophys.\/} {\bf 443}, 649 (2005).
\item[7.]\label{panagia06} N. Panagia {\it et al.}, {\it Astrophys. J.\/},
	{\bf 469}, 396 (2006).
\item[8.]\label{immler} S.I. Immler {\it et al.}, 
	{\it Astrophys. J.\/} {\bf 648}, L119 (2006). 
\item[9.]\label{hamuy} M. Hamuy {\it et al.}, {\it Nature\/} 
	{\bf 424}, 651 (2003). 
\item[10.]\label{aldering} G. Aldering {\it et al.}, {\it Astrophys. J.}, 
	{\bf 650}, 510 (2006).
\item[11.]\label{ke} The sub-luminous SN~Ia 2005ke has shown an unprecedented
	X-ray emission, which has been interpreted as the signature of
	a possible weak ejecta-CSM interaction {\it(8)}. Nevertheless, this 
        finding and the strong UV emission might be related to the nature of 
        this SN.
\item[12.]\label{benetti} S. Benetti {\it et al.}, {\it Astrophys. J.\/},
	{\bf 653}, L129 (2006).
\item[13.]\label{suzuki} S. Suzuki, M. Migliardi, 
	IAU Circ. {\bf 8667}  (2006).
\item[14.]\label{quimby} R. Quimby, P. Brown, C. Gerardy, 
	CBET {\bf 421} (2006).
\item[15.]\label{stockdale} C.J. Stockdale {\it et al.}, 
	CBET {\bf 396} (2006).
\item[16.]\label{online} Materials, Methods, Text and Figures are available 
	as supporting material on {\it Science} online.
\item[17.]\label{rand} R.J. Rand, {\it Astrophys. J.\/} 
	{\bf 109}, 2444 (1995).
\item[18.]\label{hao} H. Hao {\it et al.}, {\it Astrophys. J.} 
	{\bf 659}, L99 (2007).
\item[19.]\label{panagia07} N. Panagia, {\it Supernova 1987A: 20 Years after:
	Supernovae and Gamma-Ray Bursters}, S. Immler, R. McCray and 
	K.W. Weiler, Eds. (AIP Conf. Proc., 2007), in press; preprint available
        online (http://arxiv.org/abs/0704.1666).
\item[20.]\label{pauldrach} W.A. Pauldrach {\it et al.}, 
	{\it Astron. Astrophys.\/} {\bf 312}, 525 (1996). 
\item[21.]\label{chevalier} R.A. Chevalier, C. Fransson, 
	{\it Supernovae and Gamma-Ray Bursters},
	K. Weiler, Ed. (Lecture Notes in Physics, vol. 598, Springer Verlag, 
	New York, 2003), pp.171-194. 
\item[22.]\label{106} P.G. Judge, R.E. Stencel, {\it Astrophys. J.} 
	{\bf 371}, 357 (1991).
\item[23.]\label{xx}  I. Hachisu \& M. Kato, {\it Astrophys. J.} {\bf 558}, 
	323 (2001).
\item[24.]\label{107} L.A. Willson, {\it Ann. Rev. Astron. Astrophys.} 
	{\bf 38}, 573 (2000).
\item[25.]\label{100} I. Hachisu, M. Kato, {\it Astrophys. J.} {\bf 558}, 
	323 (2001).
\item[26.]\label{wood-vasey} W.M. Wood-Vasey,  J.L. Sokoloski, 
	{\it Astrophys. J.\/} {\bf 645}, L53 (2006).
\item[27.]\label{109} I. Hachisu, M. Kato, K. Nomoto, {\it Astrophys. J.} 
	{\bf 522}, 487 (1999).
\item[28.]\label{pp}  T.J. O'Brien {\it et al.}, {\it Nature } {\bf 442,} 
	279 (2006).
\item[29.]\label{qq}  M.F. Bode {\it et al.}, {\it Astrophys. J.}
	{\bf 652}, 629 (2006).
\item[30.]\label{bode} M.F. Bode,  {\it et al.}, {\it Astrophys. J.}, in press;
	preprint available online (http://arxiv.org/abs/0706.2745).
\item[31.] We wish to thank K. Krisciunas for providing us with the
information about the photometric evolution of SN~2006X.  We
acknowledge Katrien Steenbrugge, Francesca Primas, Michael Wood-Vasey,
Romano Corradi, Gary J. Ferland, Peter van Hoof, Fabio Bresolin and
Christopher Stockdale for useful discussions.  We particularly thank
M.F. Bode for sharing some results on RS Oph before publication. This
work is based on observations made with ESO Telescopes at Paranal
Observatory, obtained under Run IDs 276.D-5048, 277.D-5003 and
277.D-5013. Some of the data presented herein were obtained at the
W.M. Keck Observatory, which is operated as a scientific partnership
among the California Institute of Technology, the University of
California and the National Aeronautics and Space Administration. This
work made use of the Very Large Array telescope of the National Radio
Astronomy Observatory, which is operated by Associated Universities,
Inc.  under a cooperative agreement with the National Science
Foundation.
\end{enumerate}

\newpage
\normalsize

{\bf \large Supporting Online Material}

\vspace{10mm}

{\bf S1. Materials and Methods}

\vspace{5mm}

We have observed SN~2006X on 5 epochs spanning about 4 months (Table
S1). High resolution spectra were obtained with the European Southern
Observatory's (ESO) Very large Telescope (VLT) on Cerro Paranal
(Chile), equipped with the Ultraviolet and Visual Echelle Spectrograph
(UVES) {\it(S1)}, and with the Keck I Telescope equipped with the High
Resolution Echelle Spectrometer (HIRES) {\it(S2)}. For UVES we have
used the 390+580 setting, which covers simultaneously three wavelength
ranges (3290-4500\AA, 4780-5740\AA\/ and 5830-6800\AA), with a full
width half maximum (FWHM) resolution of 7 km s$^{-1}$. In the case of
HIRES, we have used a setting optimized for the NaI D region, that
covers the wavelength range 3900-8350\AA\/ with a FWHM resolution of
6.8 km s$^{-1}$. UVES data have been reduced using the UVES Data
Reduction Pipeline {\it(S3)}, while HIRES data have been processed
using standard procedures for Echelle spectra. Wavelength calibration
has been achieved using Thorium-Argon lamps. The final RMS accuracy is
about 0.15 km s$^{-1}$. The wavelength scale was corrected to the
rest-frame adopting a host galaxy recession velocity of 1571 km
s$^{-1}$ {\it(S4)}.  To compensate for the Earth's motion, a
heliocentric velocity correction has been applied to the data (Table
S1). The effect of atmospheric lines has been checked using a
spectroscopically featureless bright star (HR~3239) observed with the
same instrumental setup as for the science data.  The spectral region
of interest turns out to be essentially free of telluric features.
Therefore, the broad absorption visible starting with day +61 and
reaching $v_h\sim -$70 km s$^{-1}$ is real and not affected by
atmospheric lines. Na~I column densities have been estimated fitting
Voigt line profiles with VPFIT {\it(S5)}.  Finally, epochs have been
computed from $B$ maximum light, which took place on February 20, 2006
{\it(S6)}.

In order to set upper limits to H$\alpha$ and He~I 5876\AA\/
luminosities (Table S2), the UVES spectra have been calibrated by
means of a reference response function. Since the observations have
been obtained under sky transparency conditions that ranged from
clear/photometric (epochs $-2$, +14, and +121) to thin cirrus (+61),
taking into account the high instrument stability, the expected flux
calibration accuracy is of the order of 20-30\%.  To the best of our
knowledge, the upper limits on H$\alpha$ and He~I 5876\AA\/ presented
here are the latest ever published.  The signal-to-noise ratio on the
continuum in the Na~I D lines region ranges from $\sim$70 (day $-2$)
to $\sim$20 (day +121). 

\begin{table}
\centerline{
\begin{tabular}{ccccc}
\hline
UT Date & Phase & Instrument/Telescope & Total Integration Time &
Heliocentric Correction \\ (2006) & (days) & & (seconds) & (km
s$^{-1}$) \\
\hline
18/02    & $-$2   & UVES/VLT            & 4175      & +14.6 \\
06/03    &  +14   & UVES/VLT            & 8940      & +7.2  \\
22/04    &  +61   & UVES/VLT            & 15025     & $-$15.4   \\
06/06    &  +105  & HIRES/KECK          & 3600      & +5.0 \\
25/06    &  +121  & UVES/VLT            & 15025     & $-$28.0 \\
\hline
\end{tabular}
}
\vspace{3mm}
Table S1: High resolution spectroscopic observations of SN~2006X. 
Phase refers to $B$-band maximum light, attained on February 20, 2006
{\it(S6)}
\end{table}

To confirm the low mass of CSM estimated from the Na~I observations,
we have observed SN~2006X with the Very Large Array {\it(S8)}, in the
C configuration, at 6 cm (4.8 GHz) on November 17 and at 3.6 cm (8.4
GHz) on November 20, 2006, corresponding to +270 and +273 days after
$B$ maximum light, respectively. The SN was not detected at both
wavelengths, and the 2 SD limits are 0.07 mJy and 0.09 mJy for the two
bands respectively {\it(S9)}, which are comparable to the best limits
available so far {\it(S10)}.  At the epoch sampled by our
observations, the SN ejecta have reached a distance of $\sim$10$^{17}$
cm from the explosion site. Our upper limits on the radio flux
constrain the corresponding CSM mass to be smaller than a few
10$^{-3}$ solar masses (M$_\odot$), which is fully consistent with the
estimates deduced from the observed Na I absorptions. Very similar
upper limits have been set for standard SN Ia events like SN~1981B,
1989M, 1998bu and 1992A {\it(S10)}.

From the low resolution spectroscopic data that we have obtained in
parallel to the high resolution data set discussed in this paper, we
confirm the results of the early classification spectroscopy (S11).
SN~2006X appears to be a normal Ia, very similar to SN~2002bo
{\it(S12)}, with somewhat higher photospheric velocities. The
derivation of photometric parameters is hindered by a rather strong
reddening ($E(B-V)>$1.1) and an anomalous extinction law ($R_V<$2).
However, the decline rate, the light and color curve shapes and the
absolute magnitude are within the range of normal Type Ia SNe {\it(S6)}.

\vspace{10mm}

{\bf S2. The interstellar material}

\vspace{5mm}

The apparent position of SN~2006X is very close to a spiral arm of
NGC~4321, which is inclined by 28 degrees with respect to the line of
sight {\it(S13)}. Due to the relatively high extinction {\it(S11)},
the SN most likely exploded within or behind the disk of the host
galaxy. Moreover, the SN is projected onto the receding side of the
galaxy, and the component of the rotation velocity along the line of
sight at the apparent SN location is about 75 km s$^{-1}$ {\it(S4,
S13)}, which coincides with the strongly saturated Na~I D component
(Fig.~S1). This suggests rather unequivocally that the time-invariant,
very deep absorptions arise within the disk of NGC~4321. The
equivalent widths of the Na~I D lines, measured in the first epoch
spectrum, are 670$\pm$5 and 625$\pm$4 m\AA\/ for the D$_2$ and D$_1$
components respectively. The implied total Na~I column density,
estimated fitting multiple Voigt line profiles with VPFIT {\it(S5)},
is as large as $\log N\sim$14.3. For a Milky Way-like dust mixture
this would turn into a color excess $E_{B-V}\sim$1.1 {\it(S14)}.
Moreover, the CN lines clearly visible in our first two spectra at
$v_h$=73.6$\pm$1.0 km s$^{-1}$ (weighted average of the two epochs)
have an unprecedented depth (see also ref. {\it S15)}. On the first
epoch spectrum the equivalent widths for the three lines are:
$EW[R(0)]$=91$\pm$2, $EW[R(1)]$=60$\pm$3 and $EW[P(1)]$=42$\pm$3
m\AA\/ that, after correcting for saturation effects, correspond to
column densities of $\log [N_{R(0)}]$=14.0$\pm$0.1, $\log
[N_{R(1)}]$=13.6$\pm$0.1 and $\log [N_{R(1)}]$=13.6$\pm$0.1
respectively. Stars having comparable column densities in our own
Galaxy are known to have color excesses larger than $E_{B-V}$=1
{\it(S16)}. Applying the usual curve of growth method {\it(S16)}, an
excitation temperature of 3.0 $\pm$ 0.2 $^\circ$K is derived for this
cloud.

\begin{figure}
\centerline{
\includegraphics[width=140mm]{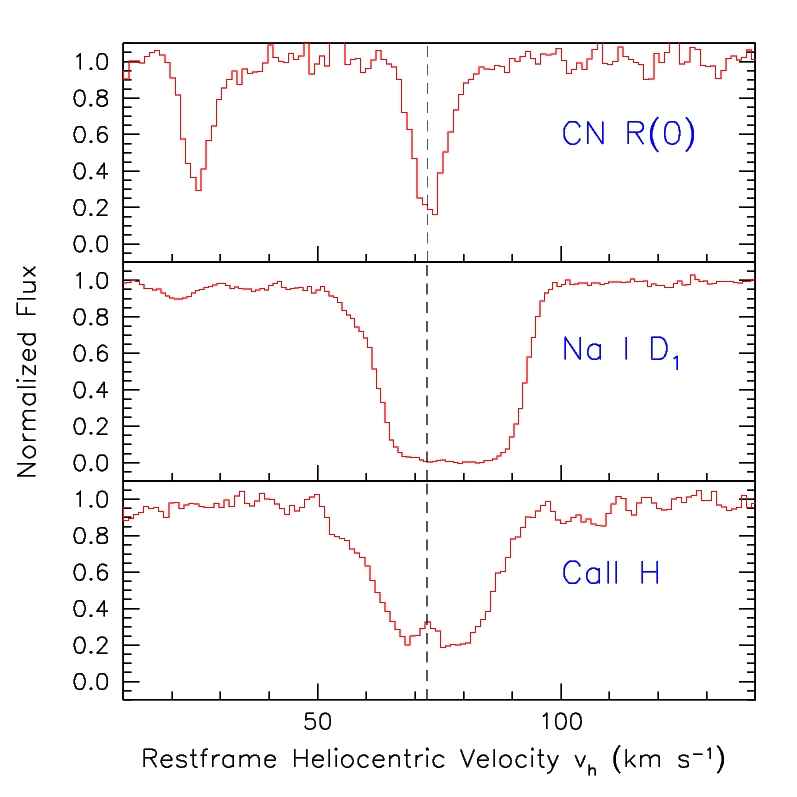}
}
\vspace{3mm}
Figure S1: \label{fig:is} Comparison between the profiles of CN R(0),
Na~I D$_1$ and Ca~II K lines for the first UVES epoch (day $-$2). The
vertical dashed line marks the velocity of the R(0) absorption center
($v_h$=73.6 km s$^{-1}$).
\end{figure}

A closer inspection of Fig.~S1 shows that the deep Na~I and Ca~II
absorptions are not due to a single cloud. In fact the Ca~II H line,
which is less affected by saturation, clearly displays at least two
components that differ in velocity by $\sim$10 km s$^{-1}$. On the
contrary, the CN lines appear as single, unresolved features (the
velocity dispersion parameter is $b\sim$2.3 km s$^{-1}$), indicating
that this molecular absorption arises within an inner subset (as seen
along the line of sight) of multiple clouds that most likely
contribute to the strong Na~I and Ca~II absorptions.

The peculiarity of the interstellar material is indicated also by the
unusual wavelength dependency of the continuum linear polarization
shown by SN~2006X. This is in fact significantly different from that
typically observed in extinguished stars within our own Galaxy
{\it(S18)}.

The presence of strongly saturated Na~ID interstellar components
could in principle mask the existence of weaker features, which
happen to fall in the same velocity range but arise in a completely
different region. Therefore, the presence of an absorption
feature at the typical velocity of a RG wind (10 km s$^{-1}$) cannot
be ruled out.  Only the study of non-extinguished objects, i.e. free
of strong interstellar absorptions, will allow us to detect slow
moving material in the surroundings of Type Ia SNe.

\begin{table}
\centerline{
\begin{tabular}{cccccc}
\hline
Phase & \multicolumn{2}{c}{H$\alpha$} & & \multicolumn{2}{c}{He~I
5876} \\
\cline{2-3} \cline{5-6}
      & 7 km s$^{-1}$ & 50 km s$^{-1}$ & & 7 km s$^{-1}$ & 50 km s$^{-1}$ \\
\hline
$-2$ & 2.2 & 16.0 & & 2.4 & 17.0 \\
+14  & 1.1 & 8.0  & & 1.1 & 7.6 \\
+61  & 0.6 & 4.0  & & 0.6 & 4.4 \\
+121 & 0.3 & 2.2  & & 0.4 & 2.6 \\
\hline
\end{tabular}
}
\vspace{3mm}
Table S2: Upper limits (3 SD) for H$\alpha$ and He~I 5876\AA\/ luminosities
measured on the UVES spectra. Luminosities are expressed in 10$^{37}$
erg s$^{-1}$ and were computed for two Gaussian line profiles with
FWHM 7 km $^{-1}$ (matching the instrumental resolution) and 50 km
s$^{-1}$, assuming a distance of 16.1 Mpc for NGC~4321 {\it(S7)}.
\end{table}

\vspace{10mm}

{\bf S3. Lack of evolution in Ca~II H\&K lines}

\vspace{5mm}

The NaI~D evolution is very clearly displayed by both D$_1$ and D$_2$
lines, which show practically identical profiles at all epochs
(Fig.~S2). Slight saturation effects are visible in the most intense
components of the D$_2$ line.

The behavior of the Ca~II H \& K lines is radically different from
that of the Na~I D lines. On the first epoch (day $-$2), the profiles
of Na~I D$_2$ and Ca~II K show very similar intensity and velocity
profiles (Fig.~S3). In particular, components A and B can be easily
identified. The differences become dramatic at the second epoch (day
+14), when Ca~II K remains practically unchanged while Na~I D$_2$
develops deep absorptions for components B, C and D. The
signal-to-noise ratio in the Ca~II region is lower, but a variation
similar to that taking place for the Na~I lines would have been
definitely detected, at least for components B, C and D. Due to the SN
fading and its spectral evolution, the continuum in the Ca~II H\&K
region becomes very weak as time goes by, making the detection of the
K line more and more difficult.  Even heavily binning our third epoch
data (day +61), the signal-to-noise ratio achieved on the resulting
profile is not sufficient to draw any firm conclusion (Fig.~S3). At
later epochs, due to the very low SN continuum, the H\&K lines could
not be detected.

\begin{figure}
\centerline{
\includegraphics[width=140mm]{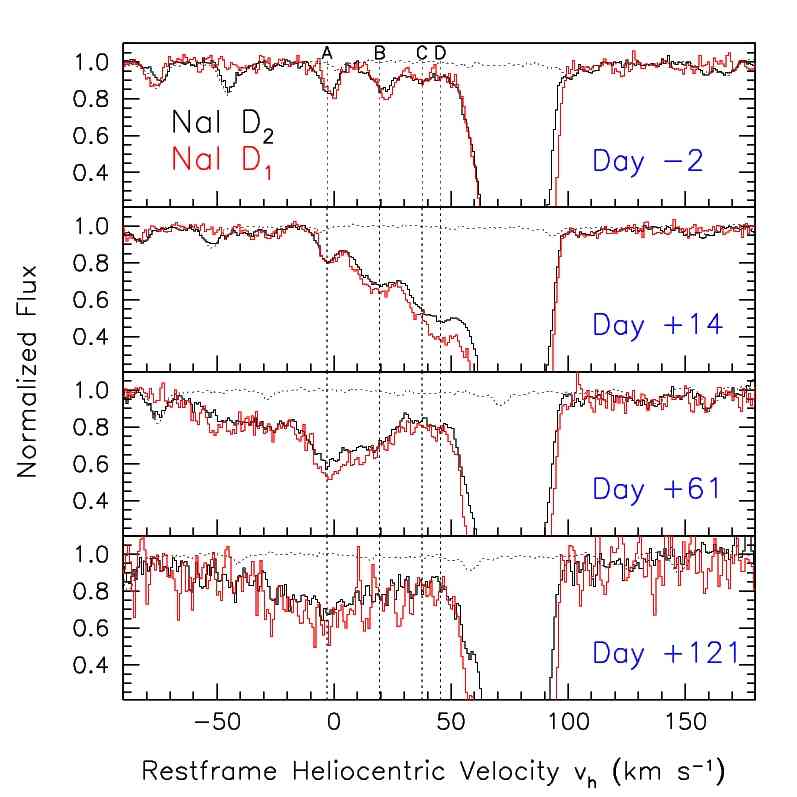}
}
\vspace{3mm}
Figure S2: \label{fig:online1} Comparison between the NaI D$_1$ and
D$_2$ line profiles on days $-$2, +14, +61 and +121. For presentation,
the intensity of the D$_1$ line has been multiplied by 2.0, that is
the D$_2$/D$_1$ ratio expected from the spin orbit statistical weights
of the Na($^2$P$_{3/2})$ and Na($^2$P$_{1/2})$ transitions. In each
panel, the dotted curve traces the atmospheric absorption spectrum in
the D$_2$ region. In order to account for the different heliocentric
corrections applied to the SN data, the telluric absorption spectrum
has been shifted in velocity to match the atmospheric features visible in
the SN spectra. The vertical dashed lines mark the main velocity components.
\end{figure}

\begin{figure}
\centerline{
\includegraphics[width=140mm]{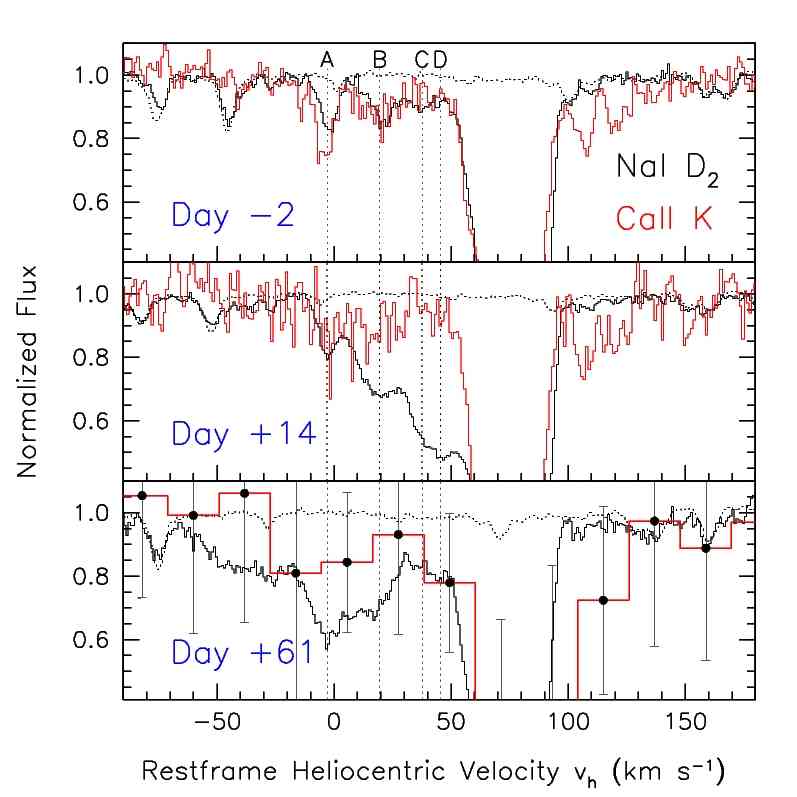}
}
\vspace{3mm}
Figure S3: Comparison between the evolution of NaI D$_2$
and Ca~II K lines during the first three epochs covered by our
observations. In order to increase the signal-to-noise ratio on day
+61 the Ca~II K data have been binned to a 0.3 \AA\/ step,
corresponding to about 23 km s$^{-1}$. The error bars indicate $\pm$1
SD level, estimated from the data within each single bin, composed by
21 original data points. In each panel, the dotted curve traces the
atmospheric absorption spectrum in the D$_2$ region.
\end{figure}

Geometric effects related to the fast expansion of the radiation ring
and the consequent change in the covering factor coupled to the patchy
structure of the intervening absorbing clouds, have been advocated to
explain the time variability claimed for some Gamma Ray Bursts (GRB)
{\it(S19, S20)}. What emerges from this analysis is that a Ca~II H\&K
evolution at the level seen in the Na~I D lines can be definitely
excluded. In turn, this rules out transverse motions or line-of-sight
effects as responsibles for the evolution observed in SN~2006X.

\vspace{10mm}

{\bf S4. The location of the absorbing material in SN~2006X}

\vspace{5mm}

After sodium is ionized by the SN radiation field, which drops quickly
after maximum light {\it(S21)}, it starts recombining with a time
scale $\tau_r$ which depends on the electron number density $n_e$ and
the radiative recombination coefficient $\beta$: $\tau_r=(n_e
\;\beta)^{-1}$. Given the time scale of the observed changes in the
NaID features, we estimate $\tau_r\sim$10 days, which implies $n_e\sim
10^5$ cm$^{-3}$ for $\beta\simeq$5$\times$10$^{-12}$ cm$^3$ s$^{-1}$
{\it(S22)}.  Such an high electron density can only come from partial
ionization of hydrogen.  Therefore, the maximum distance for the gas
where the time-variable NaI features arise is dictated by the ability
of the SN to ionize hydrogen, i.e. by the flux $S_{UV}$ of photons at
wavelengths shorter than 912\AA, times the duration $\Delta  t_{SN}$
of the UV emitting phase.

An upper limit to the number of hydrogen atoms that are ionized is
therefore given by the total number of ionizing photons, $S_{UV}\Delta
t_{SN}$, which in turn must be equal to $V_H \; n_e$, with $V_H$ being
the volume of the layer responsible for the varying absorption.
Thus, an upper limit to this volume is given by:

\begin{displaymath}
V_H\leq \frac{S_{UV}\Delta t_{SN}}{n_e}. \;\;\;\;\;\;\;\;\;\; (S1)
\end{displaymath}

Assuming that the absorbing material is
confined within a thin shell of thickness $\Delta r$ (with $\Delta r
\ll r_H $), the maximum radius of such a shell is:

\begin{displaymath} 
r_H \leq \left[ \frac{S_{UV}\; \Delta
t_{SN}}{4\pi\; n_e\; \Delta r/r} \right]^{1/3}. \;\;\;\;\;\;\;\;\;\;(S2) 
\end{displaymath}

Using $S_{UV}$= 4.4$\times$10$^{44}$ photons s$^{-1}$, the value
derived from the synthetic spectrum at maximum light {\it(S23)}, and a
light curve width of $\Delta t_{SN}$= 20 days, one gets
$r_H\leq$4$\times$10$^{15}$ cm for $n_e$=10$^5$ cm$^{-3}$ and $\Delta
r/r$=0.01.  The largest uncertainty in estimating $r_H$ comes from the
total fluence of ionizing photons ($S_{UV}\Delta t_{SN}$). In fact,
the synthetic spectrum we have used is a best fit to an HST spectrum
of SN~1992A extending down to about 1600\AA\/ and obtained 6 days past
maximum light and {\it(S24)}, i.e. when the UV flux has significantly
dropped {\it(S21)}. Nevertheless, thanks to the cubic root dependency
on it, even a 10$^3$ times higher fluence would result in an upper
limit of $\sim$5$\times$10$^{16}$ cm.  This clearly demonstrates that
the material responsible for the time-variable features is confined
well within the circum-stellar domain.

This creates a very marked distinction between Type Ia SNe and GRBs,
for some of which time-dependent UV absorption features have been
claimed {\it(S19, S20, S25)}. In fact, GRBs and their early
after-glows have a very strong X-ray/UV radiation field, so that all
possible circumstellar gas is completely ionized.  For instance,
GRB~021004 was shown to be able to completely ionize the interstellar
material out to about 100 pc ($\sim$3$\times$10$^{20}$ cm) {\it(S19)},
while for GRB~060418 the ionization was found to reach $\sim$1.7 kpc
($\sim$5$\times$10$^{21}$ cm) {\it(S25)}.

As far as calcium is concerned, we note that the flux of photons
capable of ionizing Ca~II computed from the synthetic spectrum
{\it(S23)} is $\sim$3.5$\times$10$^{46}$ photons s$^{-1}$. This is
more than four orders of magnitude smaller than the corresponding Na~I
ionizing flux.  As a consequence, the ionization fraction of Ca~II is
significantly smaller and hence, as it recombines, no detectable
effect on the depth of Ca~II H\&K lines is expected, as shown by the
observations (see S3 and Fig.~S3).  The same mechanism has been
proposed to explain the inter-stellar line variability observed in
galactic stars which, due to the much lower densities, takes place on
timescales of years to decades {\it(S26)}.

An alternative explanation for the presence of expanding shells in the
immediate surroundings of SN~2006X is that they have been lost by a
nearby star, physically not related to the progenitor system but close
enough to be influenced by the SN radiation field. In fact, shells
expanding with velocities up to several tens of km s$^{-1}$ are known
to be present, for instance, around many galactic OB associations
{\it(S26)}. Since the material has to be relatively close to
the SN in order to be influenced by its UV radiation, this alternative
scenario requires that the SN exploded within a cluster or a star
association.

In order to study the immediate surroundings of the explosion site, we
have used a deep pre-explosion image obtained with the ESO-VLT FOcal
Reducer and low dispersion Spectrograph (FORS1),
{\it(S28)}. Observations were performed on May 12, 1999 in the $R$
passband {\it(S28)}, with an exposure time of 600 seconds and a seeing
of 0.8 arcsec (FWHM), corresponding to a limiting magnitude
$R\sim$24.0 (5 SD level) at the SN location (Fig.~S4, left panel). To
better characterize the SN environment, we have searched the Hubble
Space Telescope Archive for high resolution imaging. The only suitable
data we could find were obtained with the Advanced Camera for Surveys
(ACS), {\it(S29)} on May 21, 2006 {\it(S30)} in high resolution mode
(0.027 arcsec pixel$^{-1}$) with the F435W (1480 seconds), F555W and
F775W (1080 seconds) filters. In order to increase the signal-to-noise
ratio, the three images have been registered and stacked (Fig.~S4,
right panel). The resulting image reveals a wealth of resolved
objects, which can be identified as red, blue super-giants, groups of
O stars and OB associations, with typical sizes of $\sim$100 pc. All
such associations and stellar groups (like those marked with labels A
and B on the ACS image) are clearly detected in the deep FORS1 image,
which shows also some of the brightest single stars.

\begin{figure}
\centerline{
\includegraphics[width=80mm]{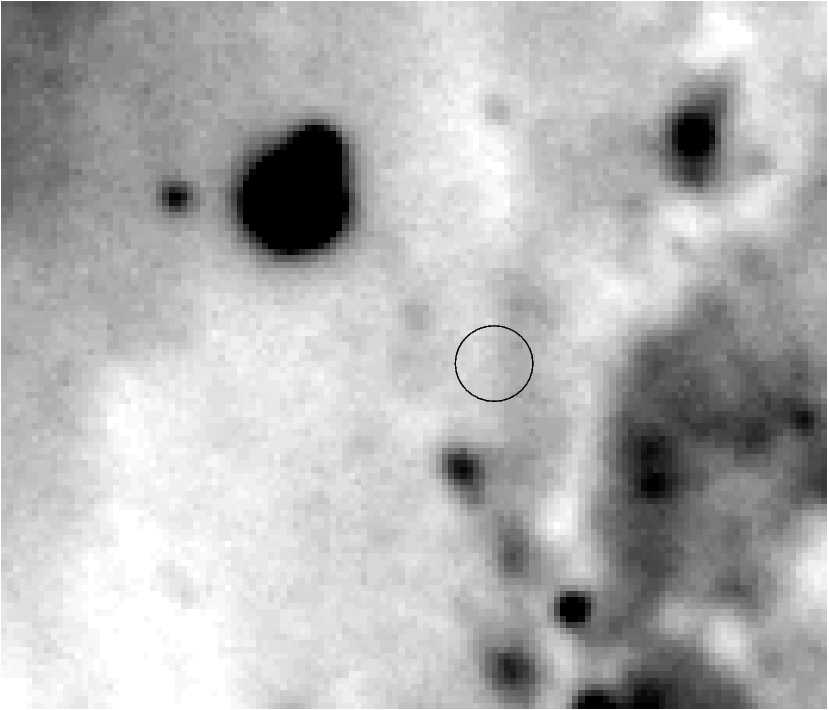}
\includegraphics[width=84.5mm]{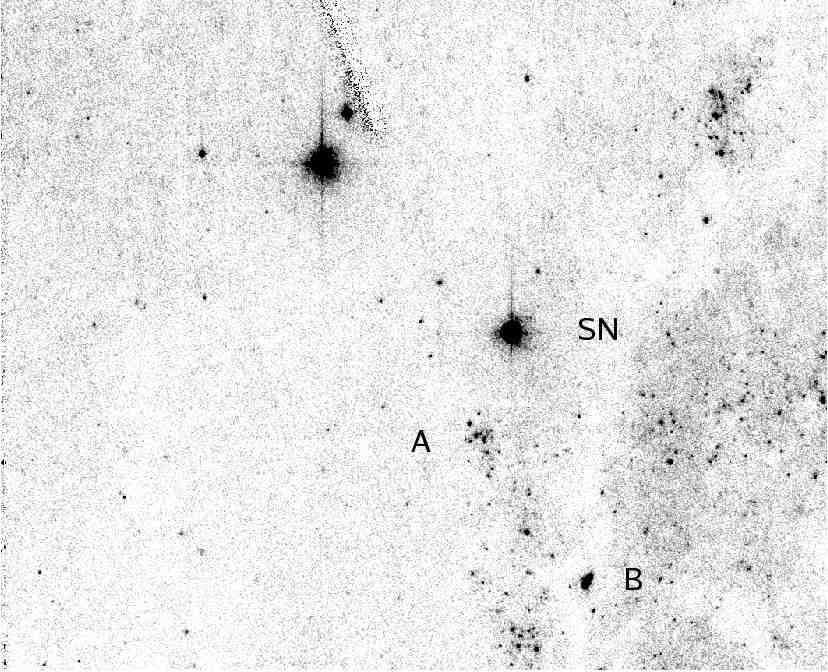}
}
\vspace{3mm}
Figure S4: \label{fig:HST}The host galaxy imaged before (left, 
VLT-FORS1) and after (right, HST-ACS) the explosion of SN~2006X.  The
field of view is 26.5$\times$22.5 arcsec for both images. The SN
position in the FORS1 image is marked by a circle. The spatial scale
on the ACS images is about 2.1 pc pixel$^{-1}$.
\end{figure}

Unfortunately, the ACS image was obtained after the explosion, when
the SN was still bright, and hence it is not possible to directly
inspect the SN surroundings. Nevertheless, the presence of groups like
that marked with A in Fig.~S4 can be excluded.  This is confirmed by
the deep FORS1 pre-explosion image, which does not show any trace of
stellar associations (like A) or even more compact groups (like B) at
the SN location. This makes the presence of an isolated star losing
material in the immediate vicinity of the SN progenitor system very
unlikely.

\vspace{10mm}

{\bf S5. Evidences of blue-shifted Na~I absorption lines in other SNe Ia}

\vspace{5mm}

Due to the large required exposure times, high resolution spectroscopy
of SN Ia is rather rare. Additionally, since the main aim of those
investigations was the study of the supposedly time-invariant
inter-stellar medium, data were usually obtained on one single epoch
around maximum light {\it(S31, S32)}. Only in recent
times, with the advent of 8-10m class telescopes, multi-epoch high
resolution spectroscopy has been performed, with the aim of detecting
possible emission lines produced by the SN ejecta-CSM interaction
{\it(S33, S34)}.  Nevertheless, due to the
underlying scientific driver of these projects, observations were
carried out during the pre- and near-maximum phase, when the Na~I D
features are at their minimum intensity in SN~2006X (Fig.~S1). For
example, the spectra of SN~2001el, observed some days before maximum
light, do not show any clear signs of Na~I D blue
components. Interestingly, they do show at least two blue components
in the Ca~II H profile at $\sim-$18 and $-$34 km s$^{-1}$ with respect
to the deep Na~I D absorption {\it(S33)}.

\begin{figure}
\centerline{
\includegraphics[width=140mm]{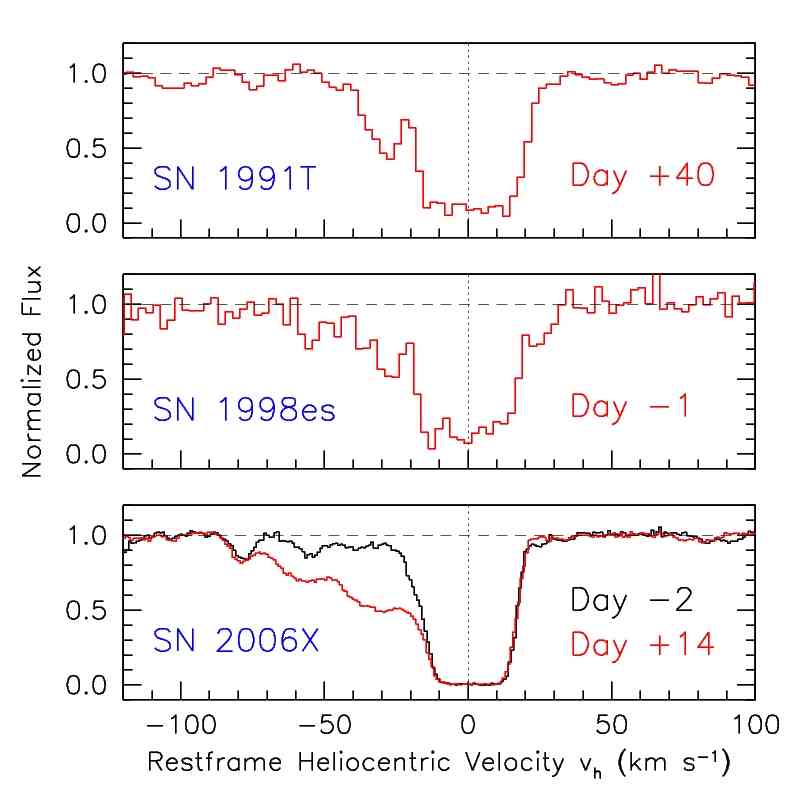}
}
\vspace{3mm}
Figure S5: \label{fig:sne}Comparison between the Sodium D$_2$ line profiles 
for three SN Ia: 1991T (upper), 1998es (middle) and 2006X (lower). For
the sake of clarity, spectra have been arbitrarily shifted in the
horizontal direction in order to place the center of the strongest
D$_2$ component at zero velocity.
\end{figure}

In the course of our long term nearby SN monitoring, we had the chance
to obtain high resolution spectroscopy of two SN Ia, namely SN~1991T
{\it(S35)} and SN~1998es {\it(S36)}. Both spectra are
unpublished. SN~1991T was observed on Jun 08, 1991 (day +40) with the
ESO 1.4m Coud\`e Auxiliary Telescope (CAT) equipped with the Coud\`e
Echelle Spectrograph {\it(S37)}, while the spectrum of SN~1998es was
obtained on Nov 24, 1998 (day $-$1) with the ESO 1.5m telescope
equipped with the Fiber-fed Extended Range Optical Spectrograph
{\it(S38)}. The Na~I D line profiles of these two objects are remarkably
similar to that of SN~2006X (Fig.~S5). Besides a strong absorption, which 
is most likely generated within the disk of the host galaxy, some blue
components are clearly visible.

To the best of our knowledge, the only other SN for which a data set
somewhat similar to that of SN~2006X has been obtained is the
core-collapse, Type IIn SN~1998S {\it(S39)}. High resolution
spectroscopy of this object has revealed a number of Na~I D components
within the host galaxy, the bluest of which ($v_h\sim-$100 km
s$^{-1}$) deepened significantly during the 19 days spanned by the two
epochs available, implying a Na~I column density increase of about 1
dex. This fact, together with the detection of time-evolving narrow H
and He P-Cyg profiles at the same Na~I velocity, was interpreted as a
signature of the outflows from the super-giant progenitor of SN~1998S,
arising in a dense shell, expanding at about 50 km s$^{-1}$
{\it(S39, S40)}.

\vspace{10mm}

{\bf S6. The CSM structure around SN~2006X: swept-up nova shells?}

\vspace{5mm}

A possible interpretation of the distinct features in the CSM is that
they arise in the remnant shells of successive novae
{\it(S41)}. The recurrent ejection of a few 10$^{-7}$ M$_\odot$ of
H at velocities of $\sim$4000 km s$^{-1}$ can create dense shells in
the slow moving material released by the companion, evacuating
significant volumes around the progenitor system
{\it(S42)}. This explanation requires that the high-velocity
nova ejecta have, by sweeping up the stellar wind of the donor star,
been slowed down to speeds several times lower than the asymptotic
shell velocities seen, for instance, in RS Ophiuchi, i.e. $\sim
300$\,km\,s$^{-1}$ {\it(S43, S44)}. For SN~2006X, if we
assume that the observed shells have all been swept up in the
energy-conserving phase {\it(S45)}, and if we take an upper limit of
the wind-mass loss rate from the companion, $\dot{M}_{\rm wind}$, as
10$^{-7}$ $M_{\odot}$ yr$^{-1}$ from the radio observations, we obtain
an upper limit on the mass ejected in each nova outburst:

\begin{displaymath}
\label{eq:Sedov}
M_{\rm nova} < 2\times 10^{-8}\,M_\odot
\frac{\dot{M}_{\rm wind}}{10^{-7} M_{\odot}\,{\rm
yr}^{-1}}  \left( 
\frac{ V_{\rm shell} }{75\,{\rm km\,s}^{-1}}\right)^2\,
\left(
\frac{ V_{\rm nova} }{1000\,{\rm km\,s}^{-1}} \right)^{-2}
\frac{ \Delta t_{\rm nova}} {10\,{\rm yr}},\;\;\;\;\;\;\;\hbox{(S3)}
\end{displaymath}

\noindent
where $M$ and $V$ refer to mass and velocity, and the subscripts
$nova$ and $shell$ refer to the mass ejected in each nova outburst and
the shell formed as the nova ejecta sweep up the wind, whilst $ \Delta
t_{\rm nova} $ represents the nova recurrence time.  

Although this value is around an order of magnitude smaller than might
be expected from recent nova calculations, we note that the ejecta
mass is a steeply decreasing quantity as the white dwarf approaches
explosion {\it{(S41, S46)}}. In addition, the estimate above assumes
that the wind from the companion is spherically symmetric. If the wind
is concentrated towards the orbital plane, and if we are observing the
shell structure close to the orbital plane, the shells in our
line-of-sight would be slowed down more than in the spherically
symmetric case. The discrepancy would be further reduced if the nova
shell is mainly slowed down in the momentum-conserving phase rather
than the energy-conserving phase, which is made more likely in this
asymmetric case where the mass is concentrated in the orbital plane.
Exactly such an axisymmetric geometry and rapid deceleration is
suggested by recent observations of RS Oph {\it(S47)}.

Indeed, the fact that there is neutral Na around the progenitor of
SN~2006X requires cool material and suggests that the shell must have
cooled down sufficiently and be in the momentum-conserving phase.

If the recurrent nova interpretation requires an aspherical geometry,
then this implies that the apparent shell structure with the
velocities we have observed would only be seen in SNe Ia which are
observed close to the orbital plane of the system. This may be a
useful future discriminant for this model. We also suggest that this
interpretation is made more satisfying if the loss of the strong Na
line at $\sim$40 km s$^{-1}$ (components C and D in Fig. 2) is due to
the interaction of the supernova ejecta with that nova shell, as that
places a useful scale on the distance of the shell from the
supernova. As a typical SN velocity of $\sim$10$^{4}$ km s$^{-1}$ is some
250 times faster than the shell, the interaction after 1 month
corresponds to a recurrence time of $\sim$25 yr, which is a plausible
value. The actual recurrence time could well be less, as that shell is
not guaranteed to be the innermost shell, merely the innermost visible
one.

As a check on the consistency of this model, we can combine our
inferred electron density of $n_{\rm e} \sim10^{5}$ cm$^{-3}$ with the
dimensions of the nova shell to produce an estimate of the mass of the
shell:

\begin{displaymath}
{\left(\frac{M_{\rm shell}}{M_{\rm \odot}}\right) \sim 1.0 \times
10^{-7} ~\left(\frac{0.1}{X}\right)~\left( \frac{\Delta
r}{10^{13}~{\rm cm}}\right)~\left( \frac{n_{\rm e}}{10^{5}~
{\rm cm}^{-3}}\right)~\left(\frac{r_{\rm shell}}{10^{15} 
~{\rm cm}}\right)^{2} },\;\;\;\;\;\;\;\;\;\;\;\;\;\hbox{(S4)}
\end{displaymath}

where, somewhat arbitrarily, we have taken a shell thickness ($\Delta
r$) that is $\sim 1\%$ of the radius of the shell. Our greatest
uncertainty is probably contained within the ionisation fraction $X$,
which is needed to obtain the mass per free electron, though the
quadratic dependence on $r_{{\rm shell}}$ alone leads to considerable
uncertainty in the estimate of $M_{\rm shell}$. Note that we have
assumed the shell is spherical; a wind concentrated in the equatorial
plane may be $\sim 10$ times less massive.

If we assume that the mass in the shell is dominated by mass swept up
from the wind (i.e. $\rm M_{\rm wind} \gg M_{\rm nova}$) and then take the
shell radius to be the product of the recurrence time and shell
velocity we can write:
\begin{displaymath}
{\left(\frac{\dot{M}_{\rm wind}}{M_{\rm \odot}/{\rm yr}}\right)~ \sim  
2.5 \times 10^{-8} 
~\left(\frac{0.1}{X}\right)~\left( \frac{\Delta 
r}{10^{13}~{\rm cm}}\right)~\left( \frac{n_{\rm e}}{10^{5}~ 
{\rm cm}^{-3}}\right)~\left(\frac{v_{\rm shell}}{50 ~{\rm km/s}}\right)^{2} 
~\left(\frac{\Delta t_{\rm nova}}{10 ~{\rm yr}}\right)
} ,\;\hbox{(S5)}
\end{displaymath}
The wind loss rate derived in this way is consistent with our
expectation in Equation~S3. 

\vspace{10mm}

\footnotesize

{\bf References}

\begin{enumerate}
\item[S1.]\label{dekker} H. Dekker {\it et al.}, {\it Proc. SPIE} 
	{\bf 4008}, 534 (2000).
\item[S2.]\label{vogt} S. Vogt {\it et al.}, {\it Proc. SPIE} 
	{\bf 2198}, 362 (1994).
\item[S3.]\label{ballester} P. Ballester {\it et al.}, {\it The Messenger} 
	{\bf 101}, 31 (2000).
\item[S4.]\label{rand2} R.J. Rand, {\it Astrophys. J.\/} 
	{\bf 109}, 2444 (1995).
\item[S5.]\label{vpfit} VPFIT has been developed by R.F. Carswell and can be
	freely downloaded\\ at
	http://www.ast.cam.ac.uk/$\sim$rfc/vpfit.html
\item[S6.]\label{krisciunas} K. Krisciunas, private communication.
\item[S7.]\label{ferrarese} L. Ferrarese {\it et al.}, {\it Astrophys. J.\/} 
	{\bf 464}, 568 (1996).
\item[S8.]\label{vla} The Very Large Array telescope of the National Radio 
	Astronomy Observatory is operated by Associated Universities, Inc.
	under a cooperative agreement with the National Science Foundation.
\item[S9.]\label{poonam} P. Chandra, R. Chevalier, F. Patat, {\it ATel} 
	{\bf 954} (2006).
\item[S10.]\label{panagia} N. Panagia {\it et al.}, {\it Astrophys. J.\/}
	{\bf 469}, 396 (2006).
\item[S11.]\label{quimby2} R. Quimby, P. Brown, C. Gerardy, 
	CBET {\bf 421} (2006).
\item[S12.]\label{benetti2} S. Benetti {\it et al.},  
	{\it Mon. Not. R. Astron. Soc.} {\bf 348}, 261 (2004).
\item[S13.]\label{canzian} B. Canzian, R.J. Allen, {\it Astrophys. J.\/}
        {\bf 479}, 723 (1997).
\item[S14.]\label{hobbs} L.M. Hobbs, {\it Astrophys. J.\/}
	{\bf 191}, 381 (1974).
\item[S15.]\label{lauroesch} J.T. Lauroesch, A.P.S. Crotts, J. Meiring, 
	P. Kulkarni, D.E. Welty, D.G. York, CBET {\bf 421} (2006).
\item[S16.]\label{crutcher} R.M. Crutcher, {\it Astrophys. J.\/} 
	{\bf 288}, 604 (1985).
\item[S17.]\label{meyer} D.M. Meyer, M. Jura, {\it Astrophys. J.\/} 
        {\bf 297}, 119 (1985).
\item[S18.]\label{wang} L. Wang, D. Baade, F. Patat, J.C. Wheeler,
	{\it CBET} {\bf 396} (2006).
\item[S19.]\label{lazzati} D. Lazzati, R. Perna, J. Flasher, V.V. Dwarkadas, 
	F. Fiore, {\it Mon. Not. R. Astron. Soc.} {\bf 372}, 1791 (2006).
\item[S20.]\label{hao2} H. Hao {\it et al.}, {\it Astrophys. J.} 
	{\bf 659}, L99 (2007).
\item[S21.]\label{panagia07b} N. Panagia, {\it Supernova 1987A: 20 Years after:
	 Supernovae and Gamma-Ray Bursters}, S. Immler, R. McCray and 
	K.W. Weiler, Eds. (AIP Conf. Proc., 2007), in press; preprint available
        online (http://arxiv.org/abs/0704.1666).
\item[S22.]\label{verner1} D.A. Verner, G.J. Ferland, K.T. Korista,
       {\it Astrophys. J.\/} {\bf 465}, 487 (1996).
\item[S23.]\label{pauldrach2} W.A. Pauldrach {\it et al.}, 
	{\it Astron. Astrophys.\/} {\bf 312}, 525 (1996). 
\item[S24.]\label{kirshner2} R.P. Kirshner {\it et al.}, {\it Astrophys. J.\/} 
	{\bf 415}, 589 (1993).
\item[S25.]\label{vreeswijk} P.M. Vreeswijk  {\it et al.}, 
	{\it Astron. Astrophys.}, in press;  preprint available
        online (http://arxiv.org/pdf/astro-ph/0611690).
\item[S26.]\label{crawford2} I.A. Crawford, {\it Mon. Not. R. Astron. Soc.}
	{\bf 328}, 1115 (2001).
\item[S27.]\label{appenzeller} I. Appenzeller {\it et al.}, {\it The Messenger}
	 {\bf 94}, 1 (1998).  
\item[S28.]\label{esoarchive} Data were obtained with ESO Telescopes at 
the Paranal Observatory under programme ID 063.H-0650(A), P.I. B. Leibundgut.
\item[S29.]\label{acs}C. Pavlovsky {\it et al.}, Advanced Camera for Surveys 
	Instrument Handbook for Cycle 16, Version 7.1 (STScI,
	Baltimore, 2006).
\item[S30.]\label{hstarchive} Images were retrieved from the data
	archive at the Space Telescope Institute. STScI is operated by
	the association of Universities for Research in Astronomy,
	Inc. under the NASA contract NAS 5-26555. Data were originally
	obtained by A. Crotts et al.
\item[S31.]\label{dodorico2} S. D'odorico  {\it et al.},
	{\it  Astron. Astrophys.\/} {\bf 215}, 21 (1989). 
\item[S32.]\label{cumming} R.J. Cumming  {\it et al.}, 
	{\it Mon. Not. R. Astron. Soc.} {\bf 283}, 1355 (1996).
\item[S33.]\label{sollerman} J. Sollerman  {\it et al.},
	{\it  Astron. Astrophys.\/} {\bf 429}, 559 (2005).
\item[S34.]\label{mattila2} S. Mattila {\it et al.},  
	{\it Astron. Astrophys.\/} {\bf 443}, 649 (2005).
\item[S35.]\label{phillips} M.M. Phillips  {\it et al.}, {\it Astron. J.}
	{\bf 103}, 1632 (1992).	
\item[S36.]\label{jha}  S. Jha, P. Garnavich, P. Challis, R.P. Kirshner,
	{\it IAU Circ.}, No. 7054 (1998).
\item[S37.]\label{enard} D. Enard, {\it Proc. SPIE} {\bf 331}, 232 (1982).
\item[S38.]\label{kaufer}A. Kaufer {\it et al.}, {\it The Messenger} 
	{\bf 95}, 8 (1999).
\item[S39.]\label{bowen} D.V. Bowen, K.C. Roth, D.M. Meyer, J.C. Blades,
	{\it Astrophys. J.\/} {\bf 536}, 225 (2000).
\item[S40.]\label{fassia} A. Fassia {\it et al.}, 
        {\it Mon. Not. R. Astron. Soc.} {\bf 325}, 907 (2001).
\item[S41.]\label{100b} I. Hachisu, M. Kato, {\it Astrophys. J.} {\bf 558}, 
	323 (2001).
\item[S42.]\label{wood-vaseyb} W.M. Wood-Vasey,  J.L. Sokoloski, 
	{\it Astrophys. J.\/} {\bf 645}, L53 (2006).
\item[S43.]\label{101b} S.R. Pottash, {\it Bull. Astr. Inst. Netherlands} 
	{\bf 19}, 227 (1967).
\item[S44.]\label{102b} R.M. Hjellming {\it et al.}, {\it Astrophys. J.} 
	{\bf 305}, L71 (1986).
\item[S45.]\label{103} L. Sedov, {\it Similarity and Dimensional Methods 
	In Mechanics} (Academic Press, New York, 1959).
\item[S46.]\label{104} O. Yaron, D. Prialnik, M.M. Shara, A. Kovetz, 
	{\it Astrophys. J.} {\bf 623}, 398 (2005).
\item[S47.]\label{105}  T.J. O'Brien {\it et al.}, {\it Nature } {\bf 442,} 
	279 (2006).
\end{enumerate} 
\end{document}